\documentclass{iau379}

\usepackage{graphicx}
\usepackage{amsmath}
\usepackage{multirow}

\begin{document}

\lefttitle{Hayashi et al.}
\righttitle{Mass profiles of dSphs in the Subaru-PFS era}

\jnlPage{1}{7}
\jnlDoiYr{2023}
\doival{10.1017/xxxxx}

\aopheadtitle{Proceedings of IAU Symposium 379}
\editors{P. Bonifacio,  M.-R. Cioni, F. Hammer, M. Pawlowski, and S. Taibi, eds.}

\title{Revealing mass distributions of dwarf spheroidal galaxies in the Subaru-PFS era}

\author{Hayashi K.$^{1,2,5}$, Dobos L.$^3$, Filion C.$^3$, Kirby E.$^4$, Chiba M.$^5$, Wyse R.$^3$,\\ \& PFS Galactic Archaeology Science Working Group}
\affiliation{$^1$ National Institute of Technology, Sendai College, Natori, Miyagi 981-1239, Japan\\
$^2$ National Institute of Technology, Ichinoseki College, Hagisho, Ichinoseki, Iwate 021-8511, Japan\\
$^3$ Department of Physics and Astronomy, The Johns Hopkins University, Baltimore, MD 21218, USA\\
$^4$ Department of Physics and Astronomy, University of Notre Dame,
225 Nieuwland Science Hall, Notre Dame, IN 46556, USA\\
$^5$ Astronomical Institute, Tohoku University, Sendai, Miyagi 980-8578, Japan\\
}

\begin{abstract}
The Galactic dwarf spheroidal galaxies (dSphs) provide valuable insight into dark matter (DM) properties and its role in galaxy formation. Their close proximity enables the measurement of line-of-sight velocities for resolved stars, which allows us to study DM halo structure. However, uncertainties in DM mass profile determination persist due to the degeneracy between DM mass density and velocity dispersion tensor anisotropy. Overcoming this requires large kinematic samples and identification of foreground contamination. With 1.25 deg$^2$ and 2394 fibers, PFS plus pre-imaging with Hyper Suprime Cam will make significant progress in this undertaking.
\end{abstract}


\maketitle

\section{Introduction}
The Galactic dwarf spheroidal~(dSph) galaxies are ideal sites for studying the basic properties of dark matter and its role in galaxy formation.
This is because these galaxies have high dynamical mass-to-light ratio $(M/L\sim 10-1000)$, which means that these are the dark matter rich systems.
Owing to their proximity of the Sun, the dSphs have the advantage that individual member stars can be resolved.
Therefore, it is possible to measure accurate line-of-sight velocities for their member stars, so that we are able to set constraints on their internal structures of dark matter halo using the high-quality data~\citep[][for reviews]{2022NatAs...6..659B,2013NewAR..57...52B}.

It is well documented that the $\Lambda$ cold dark matter ($\Lambda$CDM) theory has well-reproduced the cosmological and astrophysical observations on large spacial scales such as cosmic microwave background and large-scale structure of galaxies.
At galactic and sub-galactic scales~($\lsim 1$~Mpc), however, this concordant theory has several outstanding discrepancies between the predictions from pure dark matter simulations based on $\Lambda$CDM models and some observational facts~\citep{2017ARA&A..55..343B}.
The oldest controversial issues in $\Lambda$CDM models is the so-called ``core-cusp'' problem: dark matter halos predicted by $\Lambda$CDM simulations have strongly cusped central density profiles, whereas the dark matter halos in the observed less massive galaxies~(dSphs and low surface brightness galaxies) are suggested to have cored dark matter density profiles.
Recently, the dynamical studies for the luminous dSphs, so-called classical dSphs, have suggested that although there are still large uncertainties, these galaxies show a diversity of the inner dark matter densities~\citep[e.g.,][]{2019MNRAS.484.1401R,2020ApJ...904...45H}.
To interpret this diversity, there are various possible mechanisms such as baryonic feedback and star formation burst~\citep{2020MNRAS.497.2393L} and alternative dark matter models~\citep[e.g., self-interacting dark matter:][]{2020PhRvD.101f3009N}.

On the other hand, current dynamical studies of dSphs still face challenges in accurately measuring their central dark matter density profiles due to the well-known degeneracy between dark matter mass density and the anisotropy of the stellar velocity dispersion tensor as well as limited availability of data.
Motivated by the aforementioned problem, we, the Prime Focus Spectrograph~(PFS) Galactic Archaeology~(PFS-GA) science working group, have addressed the development of  dynamical mass models that take into account non-trivial effects such as Milky Way contamination stars, binary stars, and non-equilibrium systems.
This is because the future PFS spectroscopic surveys will enable us to improve the quantity and quality of kinematic data. Such data from PFS will open the door to placing  more stringent constraints on inner dark matter density profiles of the dSphs.

\begin{center}
  \begin{figure}
    \includegraphics[scale=.35]{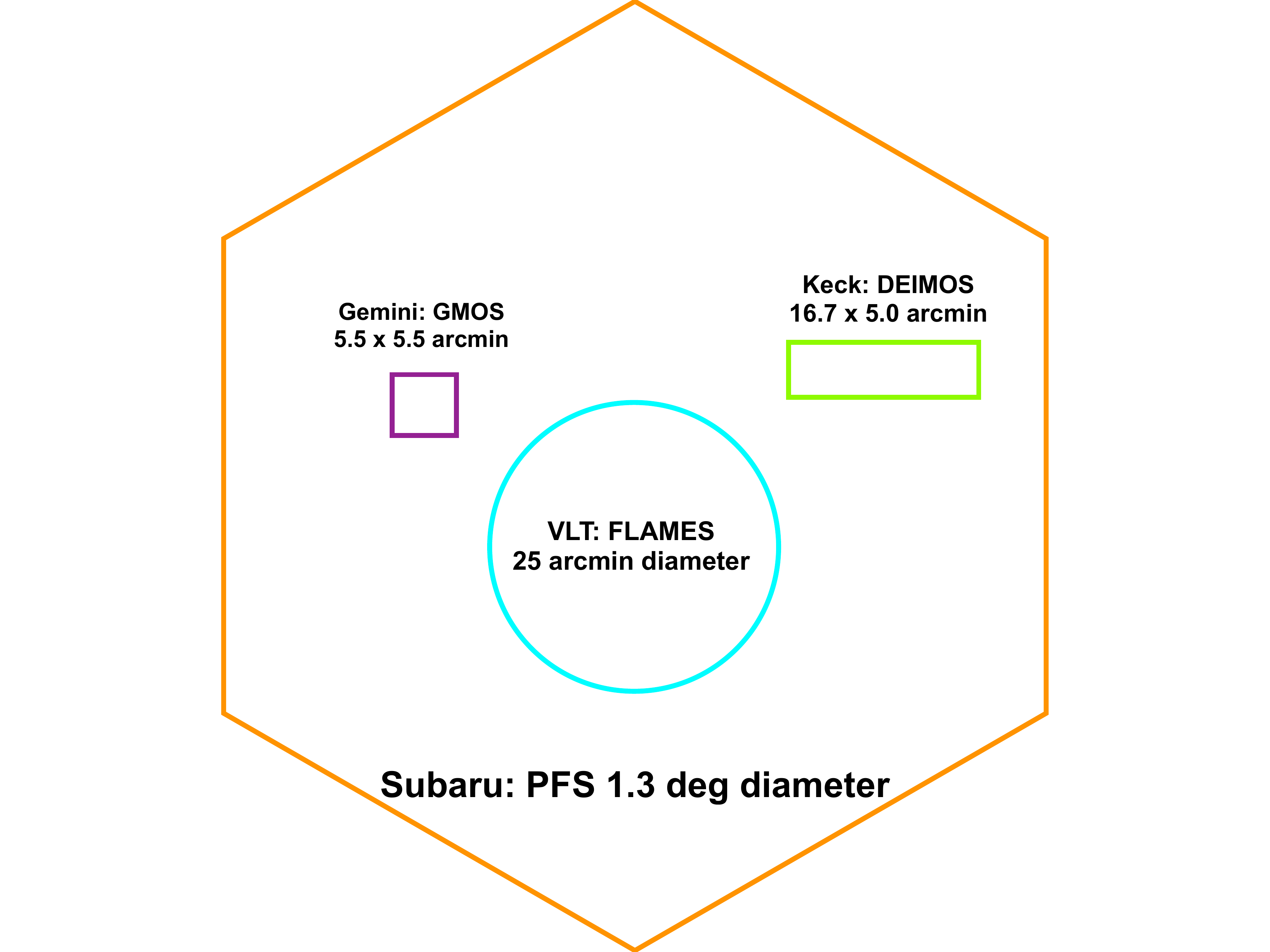}
    \caption{Schematic comparison of the field of view of Subaru PFS with that of current spectroscopy instruments such as VLT/FLAMES, Gemini/GMOS, and Keck/DEIMOS.}
    \label{fig:PFS_FoV}
  \end{figure}
\end{center}

\section{Subaru Prime Focus Spectrograph}
The Subaru Prime Focus Spectrograph attached on Subaru 8-m class telescope is a massively-multiplexed fiber-fed optical and near-infrared three-armed spectrograph~\citep{2016SPIE.9908E..1MT}.
The wave length coverage is from the optical to the near infrared ($380-1260$~nm), and thus  PFS can observe the blue and infrared ends simultaneously.
PFS is designed to allow simultaneous low and intermediate-resolution spectroscopy with blue ($380-650$~nm, R $\sim 2,300$), medium-resolution red ($710-885$~nm, R $\sim 5,000$) arm, and infrared ($940-1260$~nm, R $\sim 4,300$) arm, respectively.
Thanks to the Subaru 8-m class telescope and its Prime Focus, PFS has a large field of view (1.25 degrees in diameter in a hexagonal field) and 2400 optical fibers arranged in the hexagon. This new spectroscopy capability will allow us to obtain statistically significant samples of stellar and galactic spectra over wide areas.
Figure~\ref{fig:PFS_FoV} shows a schematic comparison of the FoV of Subaru PFS with that of VLT/FLAMES, Gemini/GMOS, and Keck/DEIMOS.
It is clear from this figure that PFS has much larger FoV than the other current spectroscopy instruments, and thus PFS will offer unique opportunities in large astronomical survey.

Taking advantage of Subaru PFS in synergy with Subaru Hyper Suprime Cam (HSC) pre-imaging, PFS-GA science working group is planning a large observation for the GA science cases in the context of a Subaru Strategic Program~\citep{2014PASJ...66R...1T}, which will start from 2024.
The main targets of our provisional plans are the Galactic dwarf galaxies, M31 disk and halo regions as well as M33 companion, and the Milky Way outer disk/halo region including stellar streams.
Here, we provide an update on the progress of the PFS-GA science project, with a particular focus on the survey of Galactic dSphs.

\begin{center}
  \begin{figure}
    \includegraphics[scale=.31]{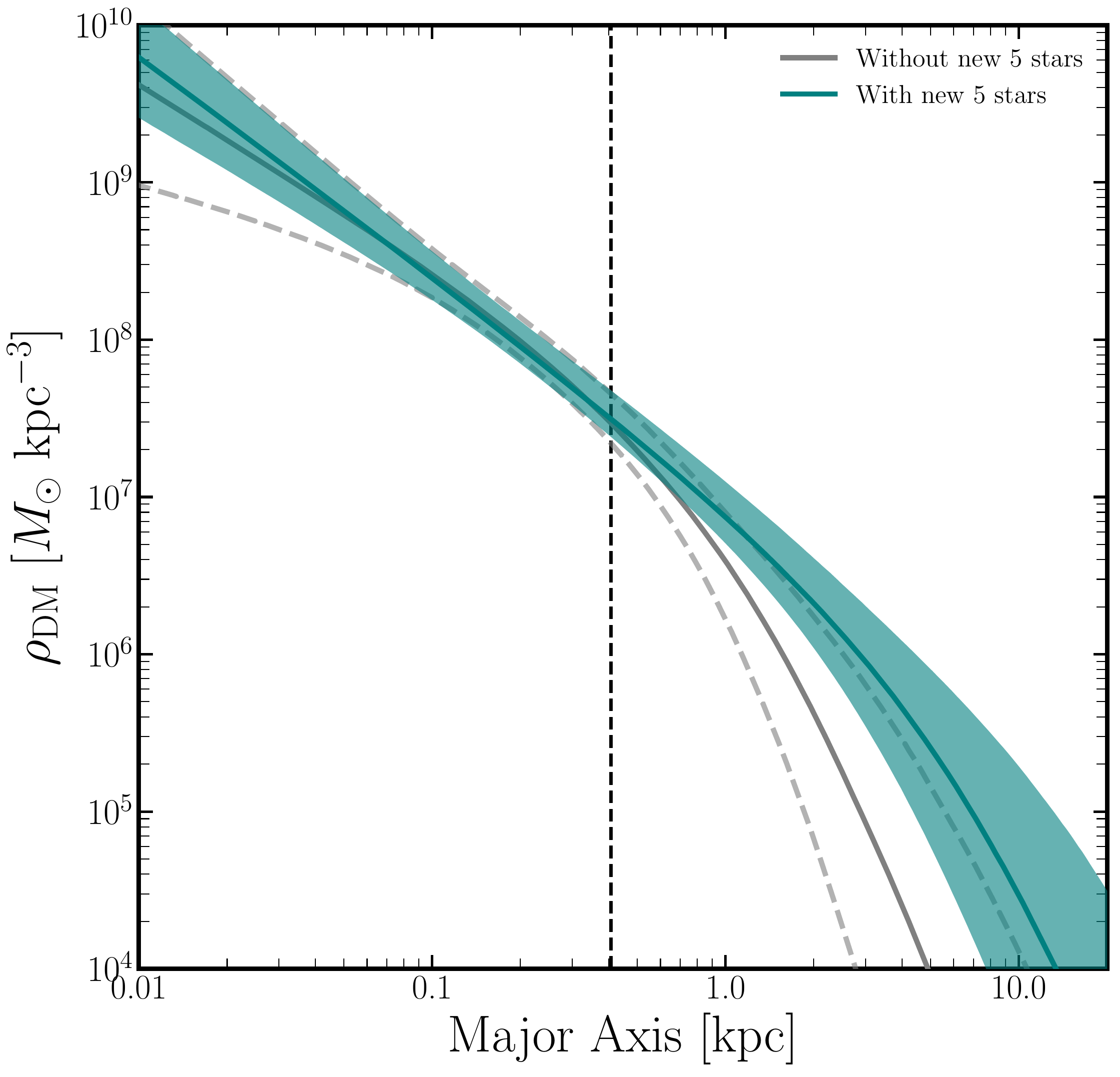}
    \includegraphics[scale=.31]{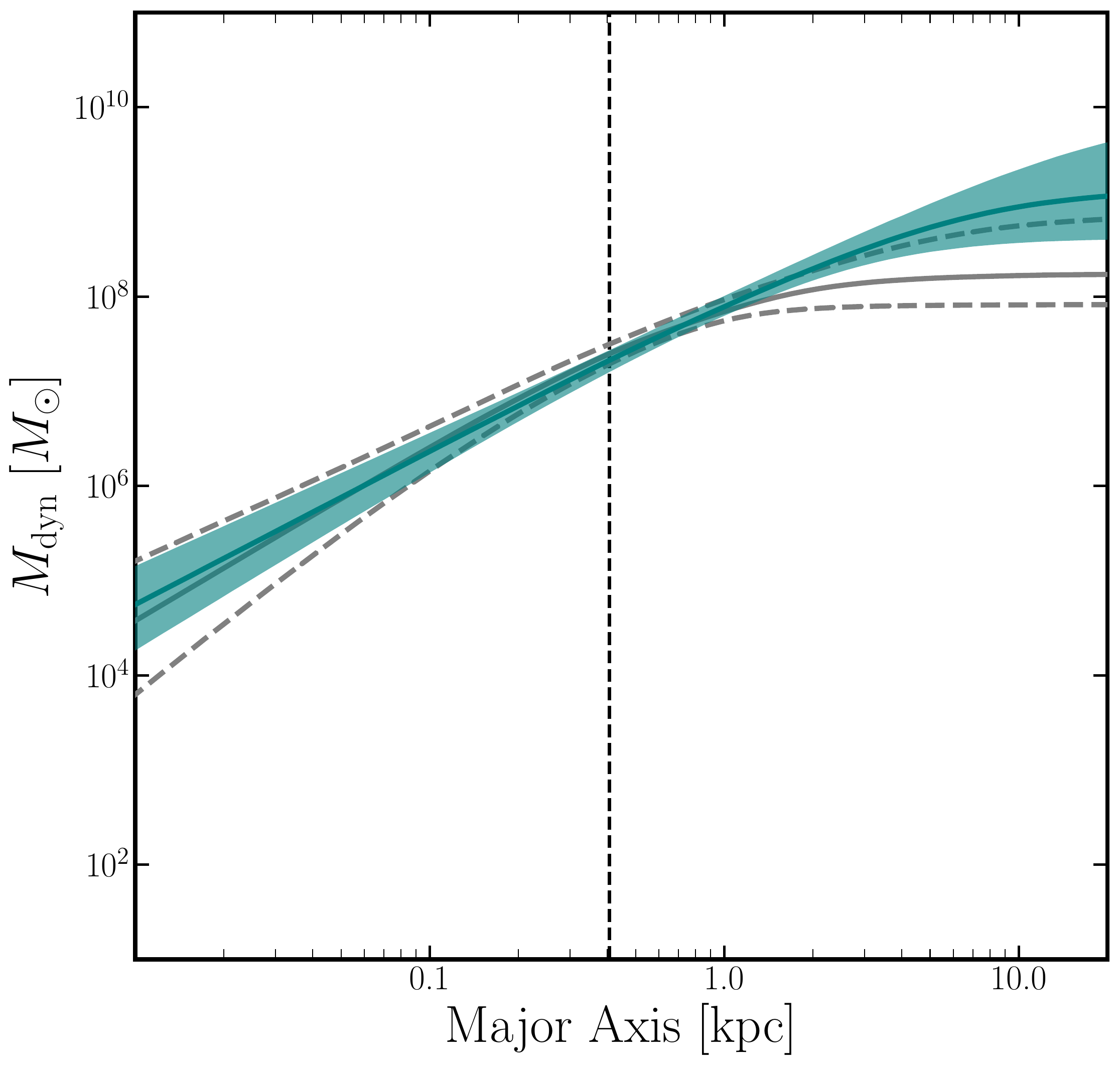}
    \caption{The estimated dark matter density profile~(left) and mass profile~(right) of Ursa Minor dwarf spheroidal galaxy. The gray solid and dashed lines are median and 68 per~cent confidence level from the kinematic sample {\it without} five new member stars~(i.e, $N_\mathrm{star}=313$), while the green solid line and shaded region are the same as gray ones but calculated from the kinematic sample {\it with} five new member stars ($N_\mathrm{star}=313+5$).}
    \label{fig:DM_UMi}
  \end{figure}
\end{center}

\section{Motivation for PFS-dSph survey}
As stated in the introduction, there are significant uncertainties in estimating the dark matter density profiles of dSphs due to the limited availability of kinematic data. The small field of view of current spectroscopy instruments has resulted in the outer regions of most dSphs being unobserved, which is one of the main reasons for this challenge.
Before presenting the current progress of the PFS-dSph survey, we demonstrate why kinematic data in the outer regions of dSphs are crucial for estimating their dark matter distributions.

Recently, new member stars of Ursa Minor (UMi), which is classified as classical dSph, are discovered by \citet{2023arXiv230113214S}.
Using Gaia proper motion and GRACES stellar spectra, they identified five bright member stars.
Surprisingly, these stars are far away from the center of Ursa Minor. 
In particular, one of them is located at 11 times half-light radius of this galaxy, which corresponds to 4~kpc as physical scale.
Adding these stars to the current available kinematic data~($N_\mathrm{star}=313$), we re-performed the axisymmetric Jeans analysis constructed by \citet{2020ApJ...904...45H}.

Here, we describe briefly our mass models as follow.
For surface stellar density distribution of a dSph, we assumed oblate Plummer profile generalized to an axisymmetric shape. For dark matter density profile, we assume a generalized Hernquist profile:
\begin{eqnarray}
&& \rho_{\rm DM}(R,z) = \rho_0 \Bigl(\frac{r}{b_{\rm halo}} \Bigr)^{-\gamma}\Bigl[1+\Bigl(\frac{r}{b_{\rm halo}} \Bigr)^{\alpha}\Bigr]^{-\frac{\beta-\gamma}{\alpha}},
 \label{DMH} \\
&& r^2=R^2+z^2/Q^2,
\label{DMH2}
\end{eqnarray}
where $\rho_0$ and $b_{\rm halo}$ are the scale density and radius, respectively; $\alpha$ is
the sharpness parameter of the transition from the inner slope $\gamma$ to the outer slope $\beta$; and $Q$ is a constant axial ratio of a dark matter halo.
These $(Q, \rho_0, b_{\rm halo}, \alpha, \beta, \gamma)$ are free parameters in our models.
Utilizing these density profiles, we solve numerically the axisymmetric Jeans equations~\citep{2008gady.book.....B} and calculate the line-of-sight velocity dispersions from these equations, where a stellar velocity anisotropy $\beta_z=1-\overline{v^2_z}/\overline{v^2_R}$ $(=\mathrm{constant})$ is taken into account.
We employ the axisymmetric mass models and Markov chain Monte Carlo techniques based on Bayesian statistics to analyze the line-of-sight velocity data of UMi dSph and obtain limits on its dark matter halo parameters.

Figure~\ref{fig:DM_UMi} shows the estimated dark matter density profile~(left) and mass profile~(right) of Ursa Minor dwarf spheroidal galaxy.
The gray lines calculated from the sample without the five new member stars (i.e., these results are the same as the ones from \citet{2020ApJ...904...45H} completely), while green ones are estimated from the sample including new member stars.
This result suggests that the addition of outermost stars can alter the dark matter density profile, although we should consider the potential influence of tidal forces on these stars. Therefore, to place robust constraints on the dark matter density profiles, it is crucial to collect a significant number of kinematic samples from stars across wide areas of the Galactic dSphs.

\begin{center}
  \begin{figure}
    \includegraphics[scale=.35]{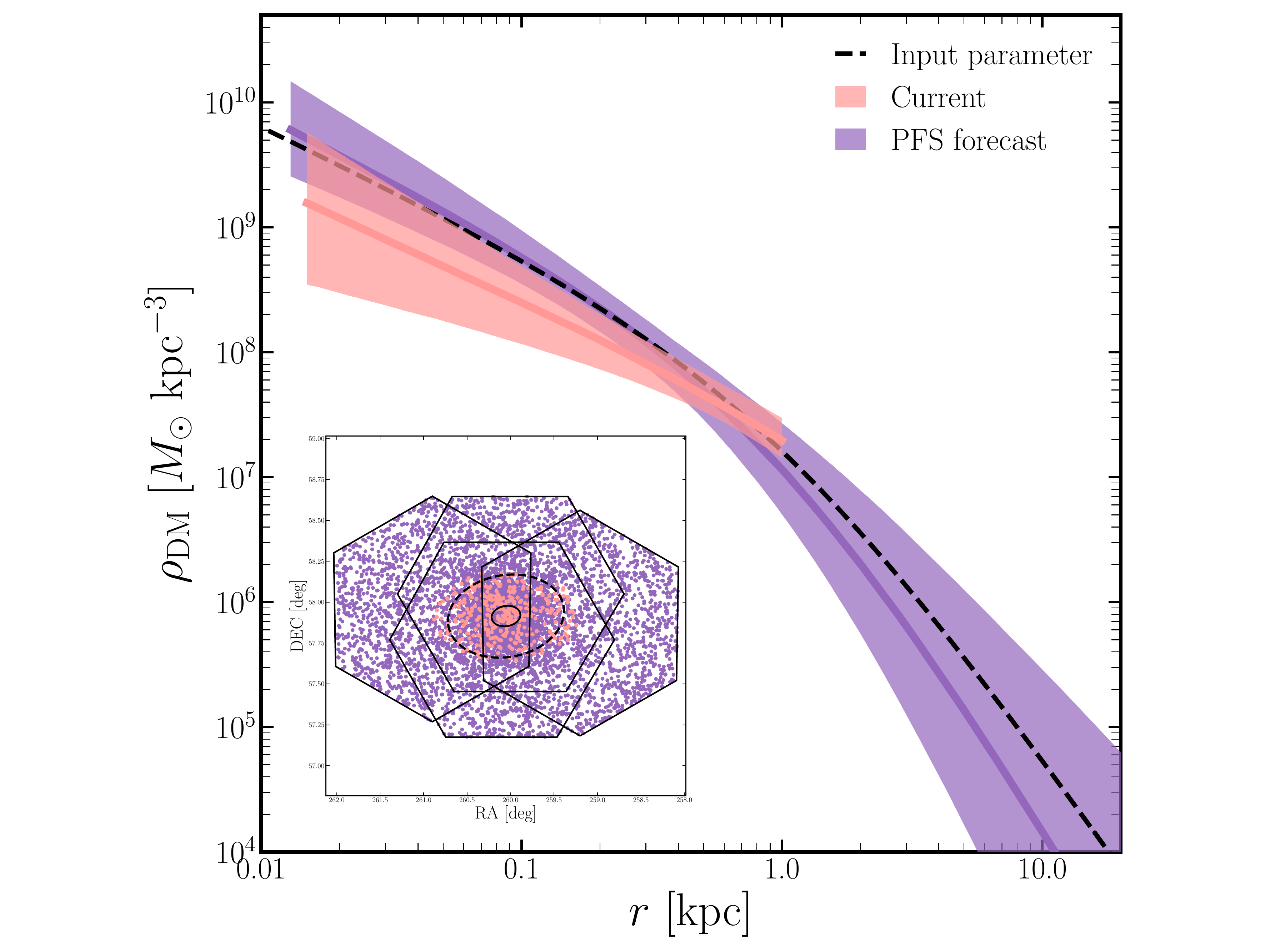}
    \caption{The dark matter density profiles derived by non-spherical mass models, using ``Current'' ($N_\mathrm{star}=500$) and ``PFS forecast'' ($N_\mathrm{star}=5000$~stars) samples shown in the inset (Draco-like mock PFS data, where solid and dashed oval lines are the half-light and nominal tidal radii, respectively). The black dashed line shows the input dark-matter density profile, whereas the colored solid lines with the shaded regions are the median values with the 68~per~cent confidence intervals obtained from each sample within the respective radial range of stars.}
    \label{fig:PFS_forecast}
  \end{figure}
\end{center}

\section{PFS forecast of dark matter density profile in a mock dSph}
To take advantage of PFS's exceptional field-of-view, depth, and spectroscopic multiplexing capabilities, we are performing mock observations and dynamical analyses for the Galactic dSphs. We here focus on Draco dSph as an example.
Mock data for dSph's member stars are generated from \texttt{AGAMA} public code~\citep{2019MNRAS.482.1525V}, which can treat non-spherical stellar and dark matter components.
Using the mock data, we conducted PFS survey simulation that was developed by the PFS science collaboration.
Thanks to the wide and deep PFS survey, we obtained approximately 5,000 stellar spectra with only four PFS pointings (see the inset in Figure~\ref{fig:PFS_forecast}).
In comparison to the current available kinematic data of Draco dSph, which consists of 500 stars, PFS will significantly improve the statistical quality of the kinematic sample.

Applying our axisymmetric mass models to the mock PFS data for Draco, we can demonstrate a significant improvement in the estimated dark matter density profile compared to the current available kinematic data. 
From Figure~\ref{fig:PFS_forecast}, the increased number of stars observed by PFS allows us to probe the kinematics of the outermost regions of Draco and thus better constrain the dark matter distribution. 
This will provide important insights into the nature of dark matter and the formation history of the Milky Way's satellite galaxies.

\section{Future plans}
It is important to keep in mind that the current mock data shown in Figure~\ref{fig:PFS_forecast} do not take into account non-negligible effects such as contamination stars, binary member stars, and dynamical non-equilibrium. These effects may affect the accuracy of the mass models and the resulting dark matter density profiles. Therefore, it is important to generate more realistic mock data and to develop more sophisticated models that can take these effects into account when analyzing the actual PFS data. 

Furthermore, in order to more precisely and accurately estimate the dark matter density profiles of the Galactic dwarf satellites, it is essential to use all of the information available in the full line-of-sight velocity distribution (LOSVD). 
This is because the LOSVD, characterized by higher-order velocity moments, is sensitive to the shape of the velocity ellipsoid, and thus should be a powerful technique for mitigating the degeneracy between dark matter density and velocity anisotropy.
Considering the shape of LOSVD can, therefore, place further constraints on the inner density slope of a dark matter halo.
Thus, for the PFS survey operation, we are developing dynamical mass models that include the LOSVD information based on Jeans analysis as well as Schwarzschild modeling, and considering the non-negligible effects using statistical approaches such as Bayesian statistics.

\section*{Acknowledgements}
This work was supported in part by the MEXT Grant-in-Aid for Scientific Research  (No.~20H01895, 21K13909, 21H05447 and 23H04009 for K.H.)


\end{document}